\documentclass[11.5pt]{article}

\usepackage[english]{babel}

\usepackage[a4paper,top=2cm,bottom=2cm,left=3cm,right=3cm,marginparwidth=1.75cm]{geometry}

\usepackage[backend=biber,style=nature]{biblatex}
\usepackage{amsmath}
\usepackage{amssymb}
\usepackage{graphicx}
\usepackage[colorlinks=true, allcolors=blue]{hyperref}
\usepackage[auth-sc, affil-it]{authblk}

\usepackage{subcaption}
\usepackage{csquotes}
\usepackage{acro}
\usepackage{siunitx}

\usepackage{booktabs}
\usepackage{colortbl}
\usepackage{xcolor}
\usepackage{xfrac}

\usepackage{caption}

\newcommand{\auth}[1]{\textit{#1}}
\newcommand{\auths}[1]{\textit{#1 et al.}}

\newcommand{\EEE}{\color{black}}

\usepackage{parskip}
\DeclareAcronym{SIFs}{
  short=SIFs,
  long=stress intensity factors,
}

\DeclareAcronym{DIC}{
  short=DIC,
  long=digital image correlation,
}

\DeclareAcronym{FEA}{
  short=FEA,
  long=finite element analysis,
}

\DeclareAcronym{ODM}{
  short=ODM,
  long=over-deterministic method,
}
\DeclareAcronym{SECT}{
  short=SECT,
  long=single-edge cracked tension,
}

\DeclareAcronym{CT}{
  short=CT,
  long=compact tension,
}

\providecommand{\keywords}[1]
{
  \small	
  \textbf{\textit{Keywords---}} #1
}

\title{Plasticity-induced crack closure identification during fatigue crack growth in AA2024-T3 by using high-resolution digital image correlation}
\author[1,*]{Florian Paysan}
\author[1]{David Melching}
\author[1]{Eric Breitbarth}
\affil[1]{German Aerospace Center (DLR), Institute of Materials Research, Linder Hoehe, 51147 Cologne, Germany.}
\affil[*]{Corresponding author: Florian.Paysan@dlr.de}

\date{\today}

\begin{document}
\maketitle


\begin{abstract}
Fatigue crack growth in ductile materials is primarily driven by the interaction between damaging and shielding mechanisms. In the Paris regime, the predominant mechanism for retardation is plasticity-induced crack closure (PICC). However, some of the mechanisms behind this phenomenon are still unclear. Identifying and separating the three-dimensional aspect from other shielding aspects during experiments is extremely complex. In this paper, we analyze the crack opening kinematics based on local crack opening displacement measurements in both 2D  high-resolution digital image correlation data and 3D finite element simulations. The results confirm that the crack opening stress intensity factor $K_\mathrm{op}$ differs along the crack path. We present a new method to determine $K_\mathrm{op}$ at the crack front allowing us to identify PICC as the predominant shielding mechanism in fatigue crack growth experiments. Furthermore, this work contributes to the discussion on the damage-reducing effect of PICC, since we find that the influence on fatigue damage in the plastic zone remains negligible when the crack is closed and crack surface contact is directed towards the surface.  
\end{abstract}

\keywords{plasticity-induced crack closure, high-resolution digital image correlation, finite element simulation}

\section{Introduction} \label{sec:intro}

From a material science perspective fatigue crack growth is mainly driven by an interaction between damaging and shielding mechanism \cite{Ritchie.1999}. While damage mechanisms occur in front of the crack tip and lead to crack propagation, shielding mechanisms have a retarding effect. \auth{Elber} \cite{Elber.1970} introduced the concept of crack closure in 1970. Under cyclic loading, the crack faces remain partially closed even though the fatigue crack is not completely unloaded and should be open. Assuming that no damage is induced in the process zone in front of the crack tip, he presents the $\Delta K_\mathrm{eff}$ concept:
\begin{equation}
	\Delta K_\mathrm{eff} = K_\mathrm{max}-K_\mathrm{op}
\end{equation}
$K_\mathrm{op}$ is declared as crack opening stress intensity factor and describes the crack tip stress, at which the crack is fully opened for the first time during loading. The crack closure is counted among the shielding mechanisms due to its damage-reducing effect according to \auth{Ritchie} \cite{Ritchie.1999}. \auth{Elber} \cite{Elber.1970} related crack surface contact to plastic deformation within the plastic wake of a growing fatigue crack. This phenomenon is also known as plasticity-induced crack closure (PICC) and is assumed to be the predominant crack closure mechanism in the Paris regime of ductile materials \cite{Chowdhury.2016}. Furthermore, it is the only crack closure mechanism that can be fully explained physically \cite{Pippan.2017}. For generic classification, the literature typically distinguishes between plane stress and plane strain conditions. \auth{Dugdale} \cite{Dugdale.1959} and \auth{Newmann} \cite{Newmann.1981} explain PICC for plane stress conditions by necking near the crack tip during crack opening. 
The material deforms plastically towards the crack surface, which is also known as out-of-plane flow. As a result, contact between the crack surfaces occurs before the crack is fully unloaded. While this explanation is widely accepted in the fracture mechanics community \cite{Pippan.2017}, the existence of PICC under plane strain conditions is debatable since, per definition, no material flow is allowed in the thickness direction \cite{Solanki.2004}. Using finite element (FE) simulations, \auth{Riemelmoser} and \auth{Pippan} \cite{Riemelmoser.1998, Pippan.2004} recognize that material flows towards the crack tip while the crack opens. In contrast to PICC under plane stress conditions, this material flow only leads to contact right behind the crack front. 
Numerical investigations by other authors validate that PICC is mostly present in plane stress conditions near the free specimen surface \cite{Camas.2020, Camas.2020b, Escalero.2021, Oplt.2019c}. 
However, 3D FE simulations are essential to investigate crack closure, but are very time-consuming due to the mesh refinement. \auths{Paysan}  performed a parameter study as a function of the load and geometry to investigate the crack contact kinematics \cite{Paysan.2022}. They conclude that the maximum stress intensity factor $K_\mathrm{max}$ and the sheet thickness $t$ are the main factors that determine the location of the crack surface contact. However, there is still a need for further research, e.g. on how 3D crack surface contact affects damage development within the plastic zone or how other mechanisms such as crack deflections or branching interact with PICC.

\begin{figure}[ht]
	\centering
	\includegraphics[width=0.90\textwidth]{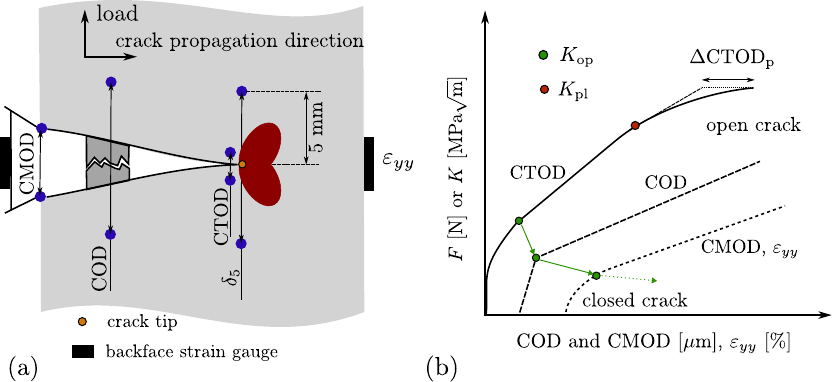}
	\caption{Summary of the current state of research towards the identification of crack closure using compliance methods: (a) measurement methods, (b) crack opening curve related to the corresponding measurement method.}
	\label{fig:crackclosure}
\end{figure}

The $\Delta K_\mathrm{eff}$ concept requires the accurate determination of  $K_\mathrm{op}$. The ASTM E647-15 standard \cite{ASTM.E647} recommends the compliance method, which assumes that $K$ is proportional to the crack opening \cite{Fleck.1991}. Backface strain or crack mouth opening displacement (CMOD) gauges are commonly used for measurement (see Figure \ref{fig:crackclosure}a). If crack closure is present, it results in a non-linear opening behavior. Both domains can be separated within the crack opening curve (see Figure \ref{fig:crackclosure}b) enabling the determination of $K_\mathrm{op}$.

Both the large distance to the crack tip position and the strong scattering in strain gauges are the main disadvantages of using compliance gauges \cite{Carboni.2007, Song.2010}. \auths{Tong} \cite{Tong.2018} claims that the study of crack closure from a systematic perspective became possible only with high-resolution digital image correlation (HR-DIC). \auth{Sutton} \cite{Sutton.1983} investigated the crack opening displacement (COD) in HR-DIC full-field data as early as 1983. \auths{Carroll} \cite{Carroll.2009} evaluate $K_\mathrm{op}$ based on COD measurements at different positions along the crack path and concluded that $K_\mathrm{op}$ strongly depends on the measurement position within the full-field displacement DIC data. Comparisons with FE simulations \cite{Alshammrei.2020, Duan.2020} confirm these findings and support the conclusions made by \auths{Carroll} \cite{Carroll.2009}. According to \auths{O'Conner} \cite{OConnor.2016} and \auths{Tong} \cite{Tong.2018b}, the value of $K_\mathrm{op}$ increases as the distance to the crack tip decreases. However, a systematic relation between $K_\mathrm{op}$ and measurement location has not yet been found. The non-linearity of the crack opening curve provides information about the crack contact \cite{Pippan.2004}. If the transition to the linear domain is smoothly curved, a continuous opening crack is assumed. If the transition is abrupt, it indicates a sudden loosening of the crack surfaces. This behavior is often associated with roughness-induced crack closure. The impact of PICC on damage evolution or crack driving force has been a topic of debate for some time \cite{Tong.2019, Kujawski.2023}. Research by \auths{Tong} \cite{Tong.2019} questions Elber's \cite{Elber.1970} claim that no damage is caused when a crack is partially closed. It was found that the $J$-integral result is only slightly different when crack closure is not taken into account. \auths{Vasco-Olmo} \cite{VascoOlmoJ.M.2019, JoseManuelVascoOlmo.2020} followed their conclusions and proposed $\Delta \mathrm{CTOD_\mathrm{p}}$ as a damage describing parameter. They found that crack tip opening displacement (CTOD) measurements are characterized by another non-linear domain in the upper region of the crack opening curve (see Figure \ref{fig:crackclosure}b). They correlate this behavior to the induction of plastic strain as a cause of the plastic zone evolution. Based on their numerical analysis, \auths{Oplt} \cite{Oplt.2023} correlated the $\Delta \mathrm{CTOD_\mathrm{p}}$ to the size of the plastic zone along the crack front. They concluded that $\Delta \mathrm{CTOD_\mathrm{p}}$ represents the plastic strain accumulation within the plastic zone region. A comparable methodology is employed by the $\delta_5$ concept, which is predominantly utilized for the investigation of fatigue cracks in thin structures \cite{Zerbst.2009}. A clip gauge is employed for the measurement of displacements in close proximity to the crack tip, with the reference points at a distance of $\SI{5}{mm}$ from each other.
However, the position the two strain gauges can remain unchanged with the advancing fatigue crack and its crack tip position. Consequently, the location dependency effect of the measurement position is not incorporated.

In the present work, we address the dependence of $K_\mathrm{op}$ on the measurement position, the influence of the 3D crack-surface contact on the damage evolution within the plastic zone, and the difficulty in identifying the dominant crack-closure mechanism in experimental data. We aim to improve the general understanding of PICC. Therefore, we use 3D FE crack propagation simulations and link them to crack opening curves based on HR-DIC data. The excellent agreement allows the investigation of the 3D contact characteristics and the associated damage accumulation within the plastic zone.

\section{Methodology} \label{sec:method}
\subsection{Specimen, Material and Loads}
\begin{figure}[ht]
	\centering
	\includegraphics[width=0.62\textwidth]{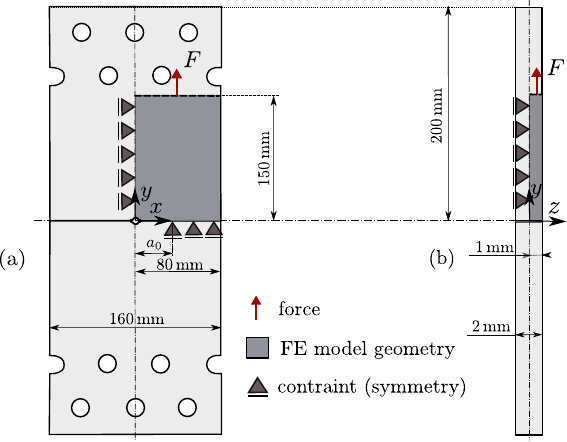}
	\caption{Dimensions of the MT(160) specimen according to ASTM E647-15 standard (highlighted in light grey) and the FE-based sub-model geometry (highlighted in dark grey) including its constraints: (a) front view and (b) side view}
	\label{fig:mt_specimen}
\end{figure}

We conducted fatigue crack growth experiments according to the ASTM 647-15 standard \cite{ASTM.E647}. A middle tension (MT) specimen with a width of $W=\SI{160}{mm}$ is used as a basis for investigating  fatigue crack growth. The dimensions of the specimen are given in Figure \ref{fig:mt_specimen}.

In this study, we consider the aluminum alloy AA2024-T3, a material with a wide range of applications in the field of aeronautics, as a reference material.
The specimen was loaded in the rolling direction, resulting in the fatigue crack growing perpendicular to the maximum grain elongation of the sheet material. The mechanical properties are given in the Table \ref{tab:properties} and taken from the literature \cite{Tamarin.2002}.

\begin{table}[htbp!]
	\centering
	\begin{tabular}{|l c c|} 
		\hline
		Elastic modulus & $E$ & 73 100\,MPa\\
		Yield stress & $R_\mathrm{p,0.2}$ & 345\,MPa\\
		Tensile stress & $R_\mathrm{m}$ & 420\,MPa\\
		Hardening modulus & $m_\mathrm{t}$ & 984\,MPa\\
		Poisson's ratio & $\nu$ & 0.33 \\
		\hline
	\end{tabular}
	\caption{Mechanical properties of AA2024-T3 used for modelling the mechanical material behavior; obtained from literature \cite{Tamarin.2002}}
	\label{tab:properties}
\end{table}

The fatigue crack starts from an initial notch with a length of $a_\mathrm{i}=\SI{8}{mm}$. We investigate mode I fatigue crack propagation at sinusoidal load with a constant amplitude in $y$-direction. The maximum force $F_\mathrm{max}=\SI{15}{kN}$ and the load ratio $R=0.1$ are constant. At a final crack length of $a_\mathrm{f}=\SI{27.8}{mm}$ we collected data to investigate fatigue crack closure, where stress intensity factors of $K_\mathrm{max}=\SI{14.9}{MPa\sqrt{m}}$ and $\Delta K=\SI{13.4}{MPa\sqrt{m}}$ are present.  

\subsection{Finite Element Simulation}
In general, the numerical simulations refer to the FE model presented in \cite{Paysan.2022} with some crucial adaptions concerning loading constraints and meshing. All FE simulations are performed with ANSYS Mechanical APDL on a RedHat Linux workstation with two Intel Xeon Gold 6240 18C CPUs and 256GB memory (DDR4-2933 RAM). For the sake of comprehensibility, a brief summary of the FE model is following given:

The FE model is based on the geometry of the MT(160) specimen, shown in Figure \ref{fig:mt_specimen}. To reduce model complexity and computational resources, we exploit symmetries of the specimen to reduce the geometry to a 1/8 model. The clamping region is excluded and all holes are neglected enabling a structured mapped meshing. Figure \ref{fig:mt_specimen} shows the constraints definition. The load $F$ is linked into a pilot node that is coupled to all nodes on the top of the model. In addition, the degrees of freedom of all nodes in the symmetry planes are restricted vertically to their corresponding plane. Within the $x-z$ plane at the bottom of the model, the initial center crack with an initial crack length of $a_\mathrm{i}$ is defined. All symmetry constraints within the interval $x \in [0, a_\mathrm{i}]$ are deleted to allow free deformation of the crack surfaces. 
The following Figure \ref{fig:meshing} sums up the most important features of the FE model.

\begin{figure}[ht]
	\centering
	\includegraphics[width=0.95\textwidth]{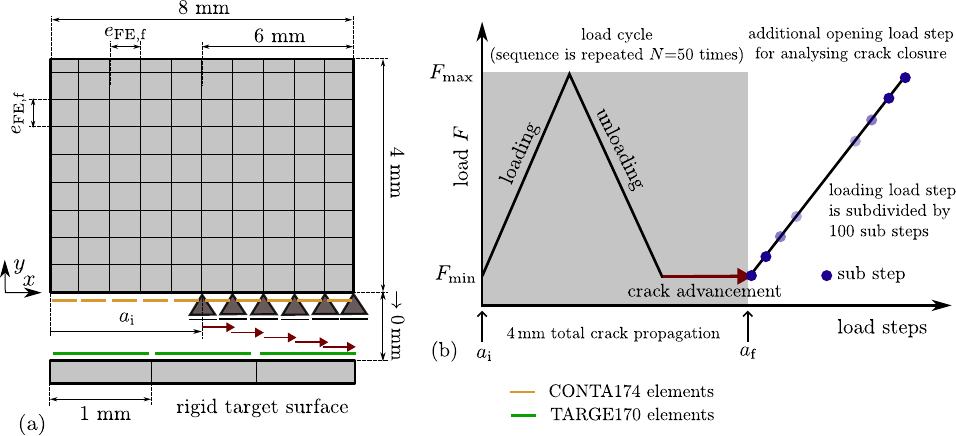}
	\caption{Characteristics of the FE model used for numerical investigation of the crack closure behavior: (a) characteristics of the refined element section, in which the artificial crack propagates according the the releasing constraint method, (b) loading sequence during the crack propagation simulation (crack propagation is conducted at minimum load), the last loading load step is separated into 100 sub steps to investigate the crack opening behaviour in detail}
	\label{fig:meshing}
\end{figure}

In general, the model is meshed using a mapped mesh strategy with 8 nodes of linear hexagonal SOLID185 elements.
The structural meshing can be divided into three separate sections: (a) the global mesh of the overall model, (b) a more refined section, schematically shown in Figure \ref{fig:meshing}a, in which the following analyses take place, and a transition section, which connects both. Within the refined section, the volume elements have an aspect ratio of 1 and an element size of $e_\mathrm{FE,f}=\SI{0.08}{mm}$. Especially contour and deformation analyses benefit from the chosen strategy, as the element shape in the observation space is always identical. However, the element size is much larger than recommended in literature \cite{Oplt.2023, Baptista.2023}. Choosing smaller element sizes results in strong deformations of single elements, which leads to convergence issues in combination with the element contact definition applied. The mesh outside the refined section is characterized by a flexible element size of $e_\mathrm{FE,g}=\SI{1}{mm}$. The contact definition is realized via contact elements. The contact is assumed to be asymmetric and all surface effects, such as crack surface roughness, are neglected. As a counterpart to the freely deformable MT(160) model, a rigid target surface in the form of a plate (SOLID185, element size $e_\mathrm{FE,c}=\SI{1}{mm}$) is defined. Following, a pairwise frictionless rigid-to-flexible contact formulation is set between the two model components. The solid elements at the crack surface of the MT(160) model are covered with CONTA174 elements and TARGE170 elements overlay the elements of the rigid contact body.  Furthermore, we use the Augmented Lagrange Algorithm as a contact solver. The contact stiffness and penetration are set to $k_\mathrm{c}=\SI{1}{N/mm^2}$ and $z_\mathrm{c}=\SI{1}{\mu m}$ with the aim of reducing the contact penetration as much as possible. The contact definition is completed with the instruction that all initial gaps are closed. The implemented model reflects the elasto-plastic material behavior using bilinear isotropic hardening and is based on the mechanical properties of AA2024-T3 given in Table \ref{tab:properties}. Anisotropic properties are neglected within the FE simulation. 

In order to analyze PICC, crack propagation needs to be performed within the elastic-plastic FE simulation. This causes a plastic wake leading to crack surface contact. We use the Releasing-Constraint \cite{Calvin.2023} method to conduct crack propagation. A single load cycle consists of a loading and unloading step followed by a node-releasing step, in which all constraints of one element row in $z$-direction are  released. Consequently, the crack advances exactly one element length $e_\mathrm{FE,f}$ within the refined mesh with each cycle at minimum load $F_\mathrm{min}$. Taking $N=50$ cycles into account, a total crack propagation of $\Delta a= a_\mathrm{f}-a_\mathrm{i}=\SI{4}{mm}$ is achieved. The application of only one loading and unloading step between crack propagation is in contrast to the general recommendation in the literature \cite{Antunes.2008, Pommier.2000} to iterate this at least twice. The authors \cite{Antunes.2008, Pommier.2000} assume that the influence of cyclic plasticity can be neglected and does not have a significant impact on the values of $K_\mathrm{op}$ as reported by \auths{Camas} \cite{Camas.2019,Camas.2020}. 

When the total crack advancement $\Delta a=\SI{4}{mm}$ and cycle number $N=50$ is reached, another load step is added, which is subdivided into 100 sub steps allowing for a detailed characterization of the crack opening behavior.

\subsection{Fatigue crack growth experiments}
\begin{figure}[ht]
	\centering
	\includegraphics[width=0.95\textwidth]{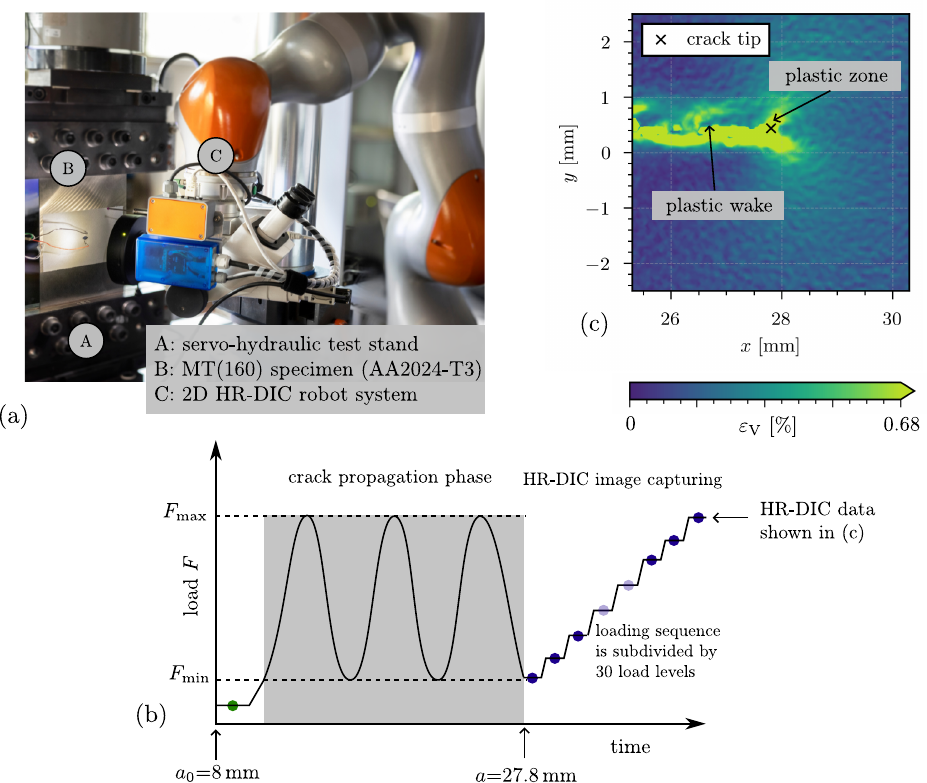}
	\caption{Experimental fatigue crack growth setup including robot-assisted HR-DIC measurement system: (a) fatigue crack growth lab setup including MT(160) specimen \cite{Paysan.2023}, (b) loading sequence and HR-DIC acquisition routine (c) HR-DIC data, von Mises strain field, at crack length $a=27.8$\,mm at $F_\mathrm{max}$}
	\label{fig:teststand}
\end{figure}

In order to compare the numerical results with experiments, we perform fatigue crack growth in MT(160) specimen made of AA2024-T3. The experimental setup is illustrated in Figure \ref{fig:teststand}, with a servo-hydraulic testing machine inducing the fatigue loading in terms of sinusoidal load with constant amplitude. Load parameters are equal to the numerical studies $F_\mathrm{max}=\SI{15}{kN}$ at $R=0.1$. For analyzing crack closure characteristics, we extended the test stand by a robot-assisted high-resolution DIC (HR-DIC) system (see Figure \ref{fig:teststand}a) \cite{Paysan.2023}. It consists of a KUKA iiwa that guides an optical Zeiss 206C stereo light microscope including a Basler a2A5320-23umPro global shutter 16 Megapixel CMOS camera. It enables HR-DIC displacement fields of the crack tip region along the entire fatigue crack propagation. A detailed description of the testing setup including the algorithms for ensuring good image quality and low DIC scattering are presented in \cite{Paysan.2023}. Figure \ref{fig:teststand}b illustrates the load sequence, including the time slots, during which the HR-DIC data acquisition is conducted.

Before the fatigue crack growth experiment starts, the robot moves the microscope to all positions on the specimen surface, where the fatigue crack is supposed to grow. The area is covered in a checkerboard manner at zero load. At each position a reference image is captured. Automatic algorithms ensure both perfect alignment with the specimen surface and good sharpness and contrast of the images. After this calibration phase, the crack propagation phase starts. The specimen is cyclically loaded until a total crack length of $a=\SI{27.8}{mm}$ is measured by a direct current potential drop (DCPD) measuring system ($I=\SI{100}{mA}, U=\SI{60}{mV}$). The system stops and the robot moves to the reference image position that best captures the crack tip area. Figure \ref{fig:teststand}c  shows the von Mises equivalent strain field. Then 30 load levels are approached, starting with the minimum load up to the maximum load. At each load level, a deformed image is captured. The comparison between the reference image and the deformed images allows to calculate the displacement and strain fields, which is done automatically using the GOM Aramis 2020 software. DIC facet size and distance are $40\,\times\,40$ pixels and $30\,\times\,30$ pixels, respectively, enabling a spatial resolution of $\SI{0.06}{mm}/\text{facet}$. 

\subsection{Crack opening displacement}
\label{sec:cod_method}
\begin{figure}[ht]
	\centering
	\includegraphics[width=\textwidth]{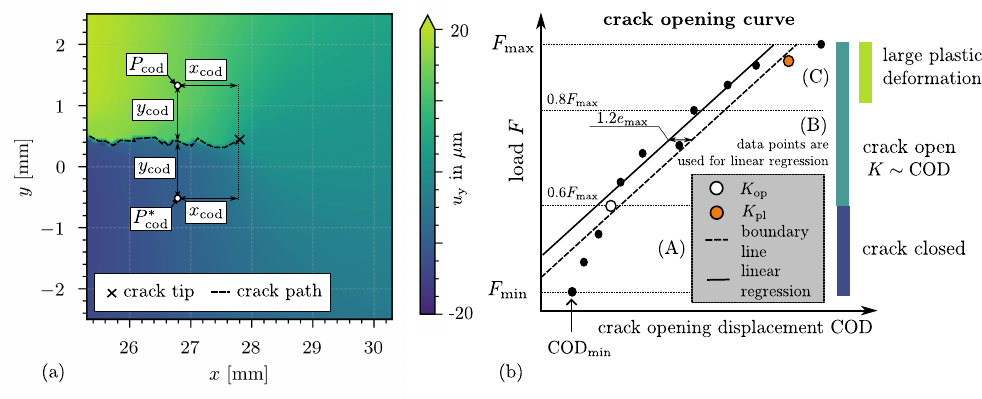}
	\caption{Determination of the crack opening stress intensity factor $K_\mathrm{op,cod}$ based on a COD measurement in HR-DIC displacement data: (a) definition of the measuring points and (b) calculation of $K_\mathrm{op,cod}$ based on the crack opening curve}
	\label{fig:cod_curve}
\end{figure}

To compare the kinematics of crack opening process of the fatigue crack, we use surface displacement data from both the FE simulation and the HR-DIC system. Therefore, the nodal solutions on the surface of the FE model are exported for this study.
We define a COD measurement location, $P_{\mathrm{cod}}$, within the two-dimensional displacement field by  surface coordinates, $P_{\mathrm{cod}} = (x_{\mathrm{cod}}, y_{\mathrm{cod}})$, where $x_{\mathrm{cod}}$ is the horizontal distance from the measurement point to the crack tip and $y_{\mathrm{cod}}$ is the vertical distance from the crack path.
Then a second measurement point, $P_{\mathrm{cod}}^* = (x_{\mathrm{cod}}, -y_{\mathrm{cod}})$, is positioned symmetrically on the opposite side of the crack path, see Figure \ref{fig:cod_curve}a. At these points the displacement component perpendicular to the crack, $u_y$, is obtained. If a measurement point does not directly coincide with a node or facet, the $u_y$ value is determined by linear interpolation from adjacent points. The local COD value is then calculated as follows:
\begin{equation}
	\text{COD} = \frac{1}{2} \cdot \big( u_y(x_{\mathrm{cod}}, y_{\mathrm{cod}}) - u_y(x_{\mathrm{cod}}, -y_{\mathrm{cod}}) \big)
\end{equation}

We plot COD against the load for each discretisation step. The resulting curve, known as the crack opening curve, is shown schematically in Figure \ref{fig:cod_curve}b.

The algorithm for the determination of the crack opening stress intensity factor, $K_{\mathrm{op}}$, from displacement data follows the methodology recommended by the ASTM E647-15 standard \cite{ASTM.E647}. The high quality of the HR-DIC displacement field data allows for a very precise analysis of crack opening. When the measurement points $P_{\mathrm{cod}}$ and $P_{\mathrm{cod}}^*$ are located near the crack tip, three distinct regions within the crack opening curve can be identified (see Figure \ref{fig:cod_curve}b):

\begin{itemize}
	\item[(A)] A non-linear relationship between load and COD is observed in the lower region, which is characteristic of crack face contact and therefore crack closure. Based on previous research \cite{Paysan.2022}, we assume that crack closure will only occur for $F < 0.6 \cdot F_{\mathrm{max}}$.
	
	\item[(B)] The middle region shows a linear relationship. The load or stress intensity factor at the crack tip is proportional to the COD, meaning that $K \sim \text{COD}$. This relationship allows the principles of linear-elastic fracture mechanics to be applied.
	
	\item[(C)] The upper region shows a non-linear relationship caused by large plastic strain accumulation within the plastic zone as reported by \auths{Vasco-Olmo} \cite{VascoOlmoJ.M.2019, JoseManuelVascoOlmo.2020}.	Based on previous research \cite{Paysan.2022}, we assume that most plastic strain occurs when loads exceed 80\% of the maximum load ($F> 0.8 \cdot F_\mathrm{max}$). 
\end{itemize}

To determine  $K_\mathrm{op}$, we use the data points from region (B) to fit a line using linear regression. The maximum deviation of a single data point in region (B) from the linear regression model is given by $e_\mathrm{max}$. That means, $e_\mathrm{max}$ is a parameter describing the inherent DIC scatter. Then, we define a boundary line by shifting the linear model by $1.2 \cdot e_\mathrm{max}$ to the right, as shown in Figure \ref{fig:cod_curve}b in order to avoid the potential for miscalculations due to the inherent HR-DIC scatter. The first data point that intersects this boundary line, starting from $F_\mathrm{min}$, is considered to be $K_\mathrm{op}$.  
In this paper, we distinguish between three different types of $K_\mathrm{op}$, further explained in Table \ref{tab:kop_types}. $K_\mathrm{pl}$ denotes the first point that leaves the linear-elastic region and crosses the boundary line again.
\clearpage
\begin{table}[htbp!]
	\centering
	\begin{tabular}{|l p{10cm}|} 
		\hline
		$K_\mathrm{op}$ & Description \\
		\hline
		$K_\mathrm{op,cod}$ & It denotes a $K_\mathrm{op}$ based on a COD measurement point\\
		$K_\mathrm{op,ctod}$ & It identifies $K_{\mathrm{op}}$ as occurring immediately beyond the crack tip.
		 Here, we consider that $x_\mathrm{cod} \to \SI{0}{mm}$. \\
		$K_\mathrm{op,cmod}$ & It characterizes $K_\mathrm{op}$ in terms similar to a CMOD measurement, as according to ASTM E647-15. It holds: $x_\mathrm{cod} = a, a_\mathrm{f}$ \\
		\hline
	\end{tabular}
	\caption{Types of $K_\mathrm{op}$ used in this paper based on COD measurements in full-field displacement data of the specimen surface}
	\label{tab:kop_types}
\end{table}

\section{Results and discussion} \label{sec:results}

We compare the numerical FE results with the  experimental  results obtained from HR-DIC measurements. If there is a satisfactory agreement on the surface, it can be assumed that the crack closure observed in the experiment will behave in a manner similar to that observed in the simulation. This enables the analysis of the 3D aspects of PICC.

\subsection{COD location dependency of $K_\mathrm{op,cod}$}
\begin{figure}[ht]
	\centering
	\includegraphics[width=0.95\textwidth]{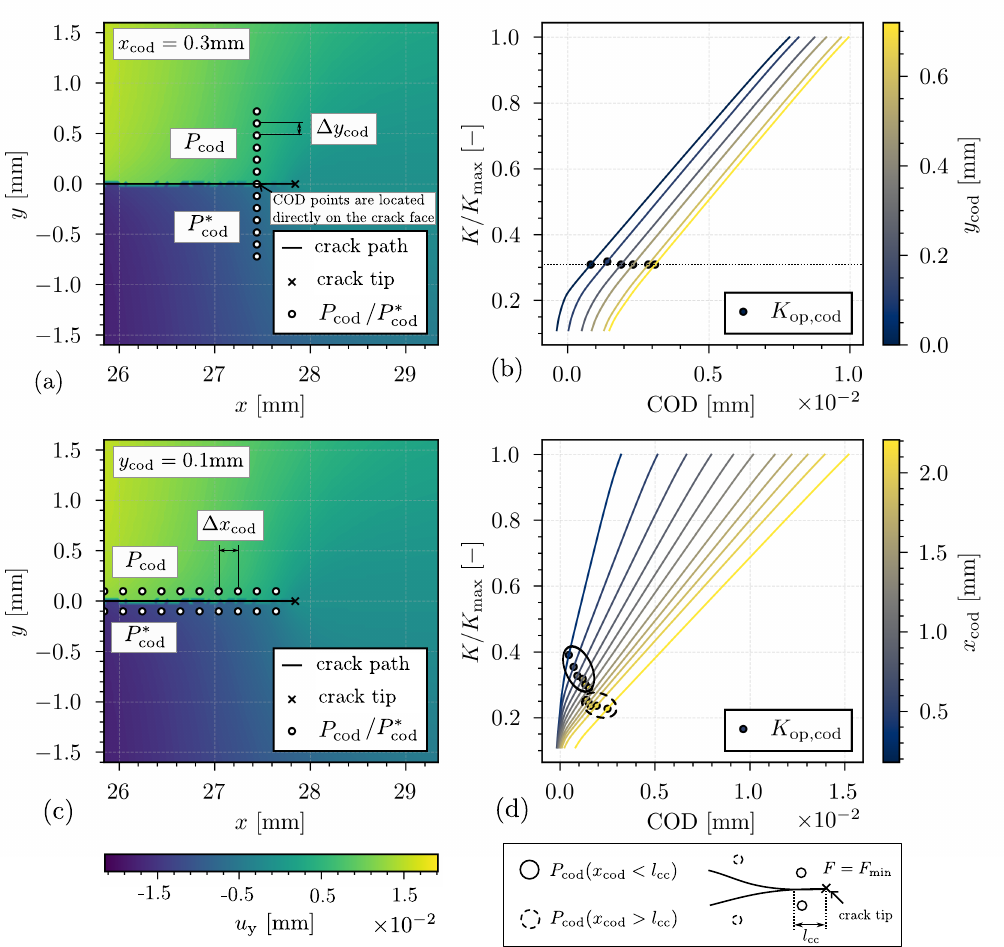}
	\caption{Results of the numerical study of the effect of $x_\mathrm{cod}$ and $y_\mathrm{cod}$ on $K_\mathrm{op,cod}$ based on FE data: $u_y$ displacement fields with the location of measurement points varying (a) $y_\mathrm{cod}$ and (c) $x_\mathrm{cod}$, (b,d) the corresponding crack opening curves including their $K_\mathrm{op,cod}$; evaluated at crack length $a=\SI{27. 8}{mm}$ with $F\in[F_\mathrm{min}, F_\mathrm{max}]$, where $F_\mathrm{min}$ and $F_\mathrm{max}$ are $1.5$ and $15\,\mathrm{kN}$, respectively.}
	\label{fig:num_results}
\end{figure}

\begin{figure}[ht]
	\centering
	\includegraphics[width=0.95\textwidth]{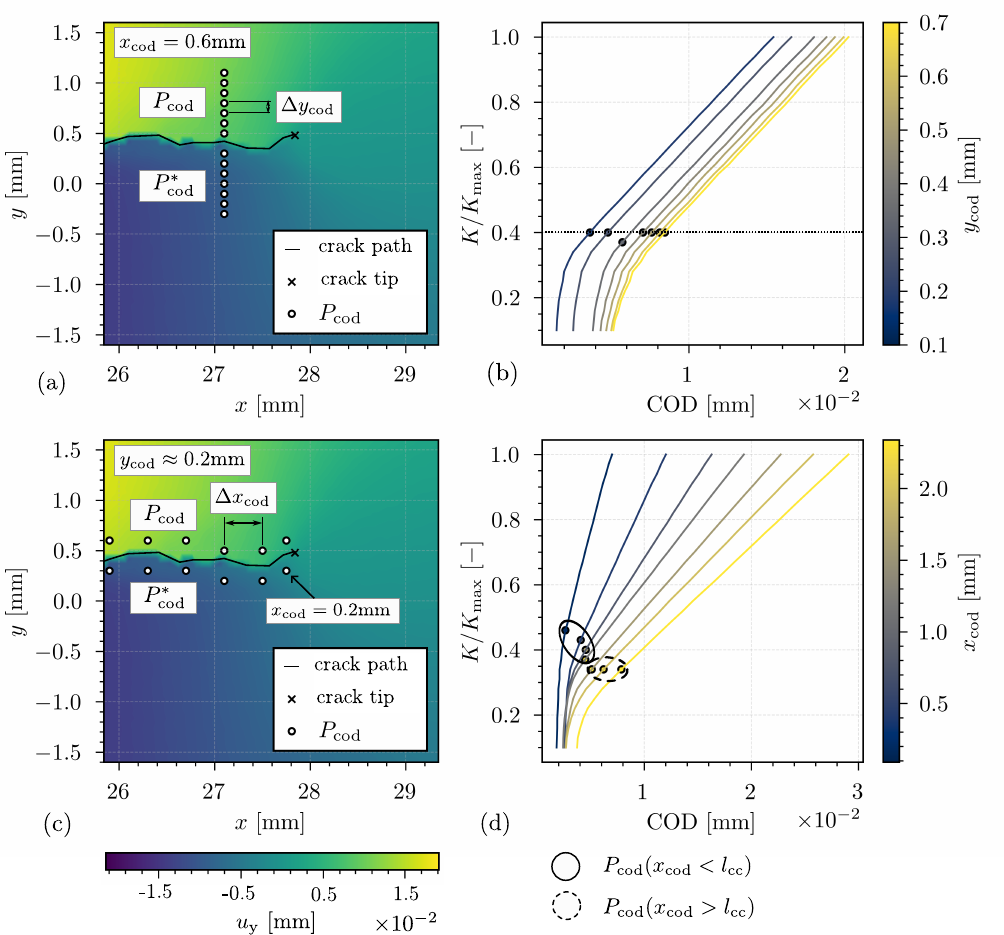}
	\caption{Results of the experimental study of the influence of the measurement location $x_\mathrm{cod}$ and $y_\mathrm{cod}$ on $K_\mathrm{op,cod}$ based on HR-DIC data: $u_y$ displacement fields with the location of measurement points varying (a) $y_\mathrm{cod}$ and (c) $x_\mathrm{cod}$, (b,d) the corresponding crack opening curves including their $K_\mathrm{op,cod}$; evaluated at crack length $a=\SI{27. 8}{mm}$ with $F\in[F_\mathrm{min}, F_\mathrm{max}]$, where $F_\mathrm{min}$ and $F_\mathrm{max}$ are $1.5$ and $15\,\mathrm{kN}$, respectively. Crack path and tip are detected by the neural networks UNetPath and ParallelNets \cite{Melching.2022}}
	\label{fig:cod_paramstudie_exp}
\end{figure}

It is widely acknowledged in the fracture mechanics community that the $K_{\mathrm{op,cod}}$ results in DIC displacement field data are sensitive to the measurement location \cite{Carroll.2009}. We investigate this dependence through a parameter study of the effects of $x_{\mathrm{cod}}$ and $y_{\mathrm{cod}}$.  The numerical results are summarized in Figure \ref{fig:num_results}, the experimental ones in Figure \ref{fig:cod_paramstudie_exp}. 

The study in Figure \ref{fig:num_results} is based on numerical surface displacement fields derived from FE analysis with a crack length of $a_{\mathrm{f}} = \SI{27.8}{mm}$. Subsequently, the crack tip is fixed to $x_\mathrm{ct} = 27.8 \, \mathrm{mm}, y_\mathrm{ct} = 0 \, \mathrm{mm}$.
In order to examine the effect of $y_{\mathrm{cod}}$, we set $x_{\mathrm{cod}} = \SI{0.3}{mm}$ to be constant. Then, we vary $y_{\mathrm{cod}}$ in a range of $y_{\mathrm{cod}} \in [0.0, 0.6]$ ($\Delta y_\mathrm{cod}=0.1$\,mm). Figure \ref{fig:num_results}a shows the location of the measurement points $P_\mathrm{cod}$ and $P_\mathrm{cod}^*$ within the $u_\mathrm{y}$ displacement field. The individual measurement points have a distance of $\Delta y_{\mathrm{cod}} = \SI{0.1}{mm}$. The crack opening curves in Figure \ref{fig:num_results}b illustrate that the non-linearity due to contact of the crack faces decreases with increasing distance $y_{\mathrm{cod}}$. In addition, it is observed that the minimum or initial crack opening displacement $\mathrm{COD_{min}}$ increases with increasing $y_{\mathrm{cod}}$. In contrast to the DIC-based studies by \auths{Sutton} \cite{Sutton.2008}, we observe no dependence on the determined $K_{\mathrm{op,cod}}$. For the examined position $x_{\mathrm{cod}} = \SI{0.3}{mm}$ we determine $K_{\mathrm{op,cod}} = \SI{4.62}{MPa\sqrt{m}} = \text{const.}$
The relationship between the crack opening curves with varying $x_{\mathrm{cod}}$ ($\Delta x_\mathrm{cod}=0.2$\,mm) is shown in Figure \ref{fig:num_results}c and d. In this case, $K_{\mathrm{op,cod}}$ depends on $x_\mathrm{cod}$. Points $P_{\mathrm{cod}}$ close to the crack tip tend to have higher $K_{\mathrm{op,cod}}$ values than those further away. In addition, the evolution of the $K_{\mathrm{op,cod}}$ values shows a gradual, ascending curvature towards the crack tip. Furthermore, the $K_{\mathrm{op,cod}}$ values shown in the crack opening curves  can be divided into two different categories:  Because of $R=0.1$ and, thus, $F_\mathrm{min} \neq 0$, the crack is only partially closed behind the crack tip due to PICC, schematically illustrated under Figure \ref{fig:num_results}d. Furthermore, it seems that especially the section of the closed crack, in which the largest contact pressure between the crack faces exists (see Figure \ref{fig:3d_picc}a), leads to a delayed crack opening near the crack tip. We denote the length of this section behind the crack tip as $l_\mathrm{cc}$. In the numerical study in Figure \ref{fig:num_results}, $l_\mathrm{cc}$ is known to be 1.36\,mm. Figure \ref{fig:num_results}d reveals that the $K_\mathrm{op,cod}$ form two different cluster depending on if they were measured at $P_\mathrm{cod}$ with $x_\mathrm{cod}<l_\mathrm{cc}$ or $x_\mathrm{cod}>l_\mathrm{cc}$. In particular, 
those encircled in black represent $K_{\mathrm{op,cod}}$ at $P_{\mathrm{cod}}$s with direct opposing crack surface contact at $F_{\mathrm{min}}$ ($x_\mathrm{cod}<l_\mathrm{cc}$), whereas the dotted highlighted ones indicate $K_{\mathrm{op,cod}}$ at initially open crack surface positions or low contact pressure between both crack faces ($x_\mathrm{cod}>l_\mathrm{cc}$). This finding leads to two conclusions: Firstly, with the use of COD measurement it is possible to identify $l_\mathrm{cc}$, if PICC is present, since the $K_\mathrm{op,cod}$ based on $P_\mathrm{cod}$ with $x_\mathrm{cod}<l_\mathrm{cc}$ behave differently than with $x_\mathrm{cod}>l_\mathrm{cc}$. Secondly, there is still a $K_\mathrm{op,cod}$ measurable although the value refers to a $P_\mathrm{cod}$ with $x_\mathrm{cod}>l_\mathrm{cc}$ . This implies the contact close behind the crack tip ($<l_\mathrm{cc}$) still influences the opening behavior of crack sections although their crack faces are not in contact at $F=F_\mathrm{min}$ or their contract pressures are so low that they rarely contribute to the delayed crack opening behavior due to PICC. 
We remark that the values of $\mathrm{COD_{min}}$ are occasionally negative. This phenomenon is caused by the FE contact definition and its allowed contact penetration.

Figure \ref{fig:cod_paramstudie_exp} shows the experimental COD measurement results from HR-DIC displacement field data at a crack length of $a = \SI{27.8}{mm}$. The crack path and the position of the crack tip are identified using the neural networks UNetPath and ParallelNets \cite{Melching.2022} implemented in CrackPy \cite{strohmann_2024_10990494}, which determined a crack tip position of $x_\mathrm{ct} = 27.8\,\mathrm{mm}, y_\mathrm{ct} = 0.36\,\mathrm{mm}$. 
In Figure \ref{fig:cod_paramstudie_exp}b we examine the effect of the vertical distance from the crack path $y_{\mathrm{cod}}$ in HR-DIC data. The results reveal no significant deviations in $K_{\mathrm{op,cod}}$ values with variations in $y_{\mathrm{cod}}$. $x_{\mathrm{cod}}=0.6$\,mm is kept to be constant. This observation confirms that $K_{\mathrm{op,cod}}$ is independent of $y_{\mathrm{cod}}$, which is consistent with the FE results shown in Figure \ref{fig:num_results}b. In addition, the non-linearity of crack closure decreases with the distance from the crack path.
The COD measurement points are located based on the detected crack path at a vertical distance of approximately $y_{\mathrm{cod}} \approx \SI{0.2}{mm}$. The horizontal distance between each COD measurement point is $\Delta x_{\mathrm{cod}}=0.4$\,mm. 
The results of the $x_{\mathrm{cod}}$ analysis in Figure \ref{fig:cod_paramstudie_exp}d are very similar to the FE results. In particular, the $K_{\mathrm{op,cod}}$ values at points close to the crack tip ($x_{\mathrm{cod}} < \SI{1.4}{mm}$) are comparable to the FE results, showing the ascending characteristic. In the following, the analysis given indicates that at minimum load $F=F_\mathrm{min}$ the crack is closed to $l_\mathrm{cc}=1.4$\,mm. For $x_{\mathrm{cod}} > \SI{1.4}{mm}$ the $K_{\mathrm{op,cod}}$ values are almost constant.

In conclusion, our study highlights the negligible influence of $y_{\mathrm{cod}}$ on $K_{\mathrm{op,cod}}$. Deviations from the results of previous studies \cite{Duan.2020, Carroll.2009} can likely be attributed to the applied measurement methods. Increasing the distance $y_{\mathrm{cod}}$ results in a flattening of the crack opening curves, and thus increases the effect of measurement noise in HR-DIC data on the determination of $K_{\mathrm{op,cod}}$. For an accurate characterization of crack closure behavior using HR-DIC, we advise to minimize the noise in DIC measurements and ensure high temporal resolution of the crack opening process. Consequently, despite the negligible influence of $y_{\mathrm{cod}}$, we recommend to aim for the smallest possible vertical distance of the measurement point $P_{\mathrm{cod}}$ to the crack path. This allows a robust segmentation of the non-linear from the proportional part within the crack opening curve and thus a stable identification of the $K_{\mathrm{op,cod}}$ value. However, both results, numerical and experimental ones, confirm the spatial dependency of $K_{\mathrm{op,cod}}$ along the crack path \cite{Carroll.2009, Alshammrei.2020, Duan.2020} and supports the hypothesis of an increasing $K_{\mathrm{op,cod}}$ with decreasing distance to the crack tip $x_{\mathrm{cod}}$ \cite{OConnor.2016,Tong.2018b}.

\subsection{Identifying plasticity-induced crack closure}
As crack surface roughness is not considered, our FE crack propagation simulation only covers the effect of plasticity induced crack closure. Based on the agreement of the $K_\mathrm{op,cod}$ behavior in our numerical and experimental studies in Figure \ref{fig:num_results}d and \ref{fig:cod_paramstudie_exp}d, we conclude that plasticity induced crack closure is the dominant crack closure mechanism here (AA2024-T3; $\Delta K = \SI{13.4}{MPa\sqrt{m}}$). 

\begin{figure}[ht]
	\centering
	\includegraphics[width=0.95\textwidth]{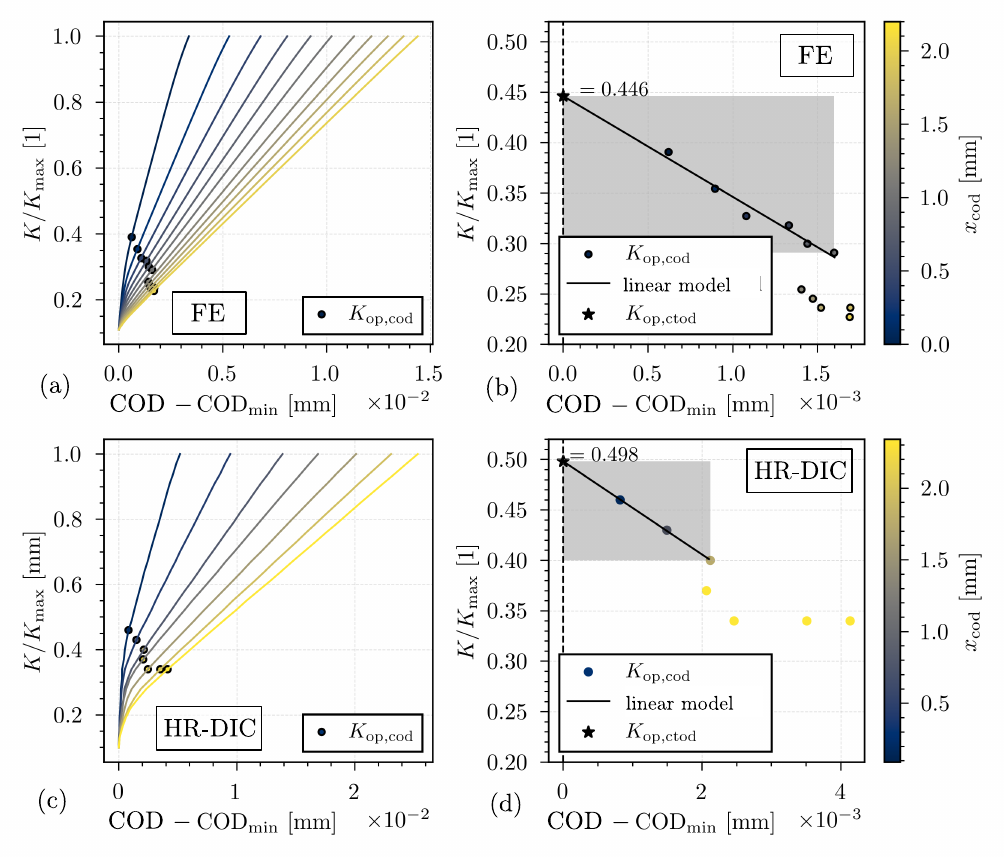}
	\caption{ Our method for determining $K_\mathrm{op,ctod}$ at the crack tip based on COD measurement points along the crack path: crack opening curves shifted to a common minimum based on (a) FE and (c) on HR-DIC data, linear relation of $K_\mathrm{op,cod}$ enabling the estimation of $K_\mathrm{op,ctod}$ - applied in (b) FE und (d) HR-DIC data}
	\label{fig:picc_kriterium}
\end{figure}

In the following, we  aim to  define a criterion to identify PICC that bases on the $K_\mathrm{op,cod}$ dependence along the crack path. First, we only take into account for the $K_\mathrm{op,cod}$ that bases on $P_\mathrm{cod}$ with $x_\mathrm{cod} < l_\mathrm{cc}$. Afterwards, we correct the crack opening curves around their initial COD value ($\text{COD} - \text{COD}_{\mathrm{min}}$). This is permissible, since we know on the one hand, that contact penetration as allowed within the FE simulation but does not occur in reality. On the other hand, we showed that $y_\mathrm{cod}$, the second reason for a possible horizontal shift of the crack opening curve, does not affect the resulting $K_\mathrm{op,cod}$. Figure \ref{fig:picc_kriterium}a and c show the crack opening curves of FE and HR-DIC data shifted to a common minimum. Figures \ref{fig:picc_kriterium}b and d present the  $K_{\mathrm{op,cod}}$ values separately. 

When the COD values of the crack opening curves are aligned to a common minimum, the $K_{\mathrm{op,cod}}$ values exhibit a linear characteristic when related to $P_\mathrm{cod}$ with $x_\mathrm{cod} < l_\mathrm{cc}$ (highlighted in grey in Figures \ref{fig:picc_kriterium}b and d). Physically, this represents a uniform crack opening towards the crack tip, which is assumed to be a typical characteristic of the crack opening kinematics in presence of PICC \cite{Pippan.2017}. Thus, we define the following criterion:

\textbf{Criterion for plasticity-induced crack closure}\\
\textit{Let $y_{\mathrm{cod}}$ be fixed and $x_{\mathrm{cod},i}, i=1,\dots,n$, be $x$-values of the COD measurement points along the crack path. Let $K_{\mathrm{op},i}$ be the crack opening stress intensity factor determined for the COD curve at measurement point $(x_{\mathrm{cod},i}, y_{\mathrm{cod}})$, and $\lambda_1,...,\lambda_{n}$ be the minimum-shifted crack opening displacements $\lambda_i := \text{COD}(x_{\mathrm{cod},i},y_{\mathrm{cod}}) - \text{COD}_{\mathrm{min}}$.
PICC is present and the pre-dominant crack closure mechanism at measurement points $I \subset \{1,\dots,n\}$, if the following condition holds true: There exists a linear fit, represented by slope and intercept $c,k \in \mathbb{R}$, such that for all $i \in I$:
\begin{equation*}
	\label{eq:picc_criterion}
	|\lambda_i - c \cdot K_{\mathrm{op},i} - k|<\varepsilon_{\mathrm{picc}},
\end{equation*}
Here, $\varepsilon_{\mathrm{picc}} > 0$ defines the permissible deviation of the points from the linear approximation.}

Loosely speaking, PICC is present where $K_{\mathrm{op,cod}} \propto \text{COD}-\text{COD}_{\mathrm{min}}$ with an allowable deviation of $\varepsilon_{\mathrm{picc}}$. This threshold should depend on the discretisation of the crack opening process to reflect the idea that there is less noise in the determination of $K_{\mathrm{op,cod}}$ if the discretization of COD curves is finer, we define $\varepsilon_{\mathrm{picc}} = \Delta K/N_{\mathrm{L}} \cdot s$. Here, $N_{\mathrm{L}}$ denotes the number of load steps or sub steps, respectively, used for the discretisation of the crack opening process and $\Delta K = K_{\mathrm{max}}-K_{\mathrm{min}}$. The factor $s > 0$ is adjustable and can be used to potentially make the criterion more robust. However, it is crucial that the linear character of the $K_{\mathrm{op,cod}}$ values, as illustrated in Figure \ref{fig:picc_kriterium}, is given. We recommend to keep $s$ as small as possible.
Assuming that the crack continues to open uniformly up to the crack tip, the crack tip opening load $K_{\mathrm{op,ctod}}$ can be estimated by setting the slope of the linear model to zero $(c=0)$ in Equation \ref{eq:picc_criterion}. It follows:
\begin{equation}
	K_{\mathrm{op,ctod}} \approx k
\end{equation}

Table \ref{tab:cod_comparison_fe_dic} compares the determined values for applying the criterion based on FE analysis with those obtained from HR-DIC data.
\begin{table}[h]
	\centering
	\begin{tabular}{|lcc|}
		\hline
		Criterion Parameter & FE Data & HR-DIC Data\\
		\hline 
		$K_\mathrm{max}$ $\mathrm{[MPa\sqrt{m}]}$ & 14.9 & 14.9\\
		$\Delta K$ $\mathrm{[MPa\sqrt{m}]}$ & 13.4 & 13.4\\
		$N_{\mathrm{L}}$ $\mathrm{[1]}$ & 100  & 30\\
		$s$ $\mathrm{[1]}$ & 2 & 1\\
		$c$ $\mathrm{[MPa\sqrt{m}/\mu m]}$ & -1.35  & -1.38\\
		$k$ $\mathrm{[MPa\sqrt{m}]}$ & 6.63  & 7.42\\
		$\varepsilon_{\mathrm{picc}}$ $\mathrm{[MPa\sqrt{m}/\mu m]}$ & 0.13 & 0.45\\
		$K_{\mathrm{op,ctod}}$ $\mathrm{[MPa\sqrt{m}]}$ & 6.63 & 7.42\\
		\hline
	\end{tabular}
	\caption{The application of the criterion for the presence of PICC on FE and HR-DIC data.}
	\label{tab:cod_comparison_fe_dic}
\end{table}
Comparing the $K_{\mathrm{op,ctod}}$ result with the sub-step in the FE crack growth simulation when there is initially no contact during crack opening across the crack faces, $K_{\mathrm{op,FE}} = 0.43 \cdot K_{\mathrm{max}} = 6.47 \, \mathrm{MPa\sqrt{m}}$, results in a deviation of less than 3\,\%. The $K_{\mathrm{op,ctod}}$ result based on HR-DIC data is larger compared to the FE data, which can be attributed to significantly higher contact pressures in real fatigue cracks. The high incremental crack advancement in the FE crack growth simulation reduces plastic strain accumulation within the plastic zone, subsequently leading to lower contact pressures on the crack surface. In actual DIC data, measuring crack opening this close behind the crack tip is not feasible due to generally insufficient local resolution of the measurement. However, this method still allows for the determination of $K_{\mathrm{op,ctod}}$. Generally, the experimental determination of this value is challenging, and a reliable procedure has not been found so far \cite{Pippan.2017}. This methodology provides a potential approach to address this issue and thus improve the understanding of crack opening kinematics of plasticity induced crack closure in experimental data.

\subsection{3D aspects of plasticity-induced crack closure}
\begin{figure}[ht]
	\centering
	\includegraphics[width=0.9\textwidth]{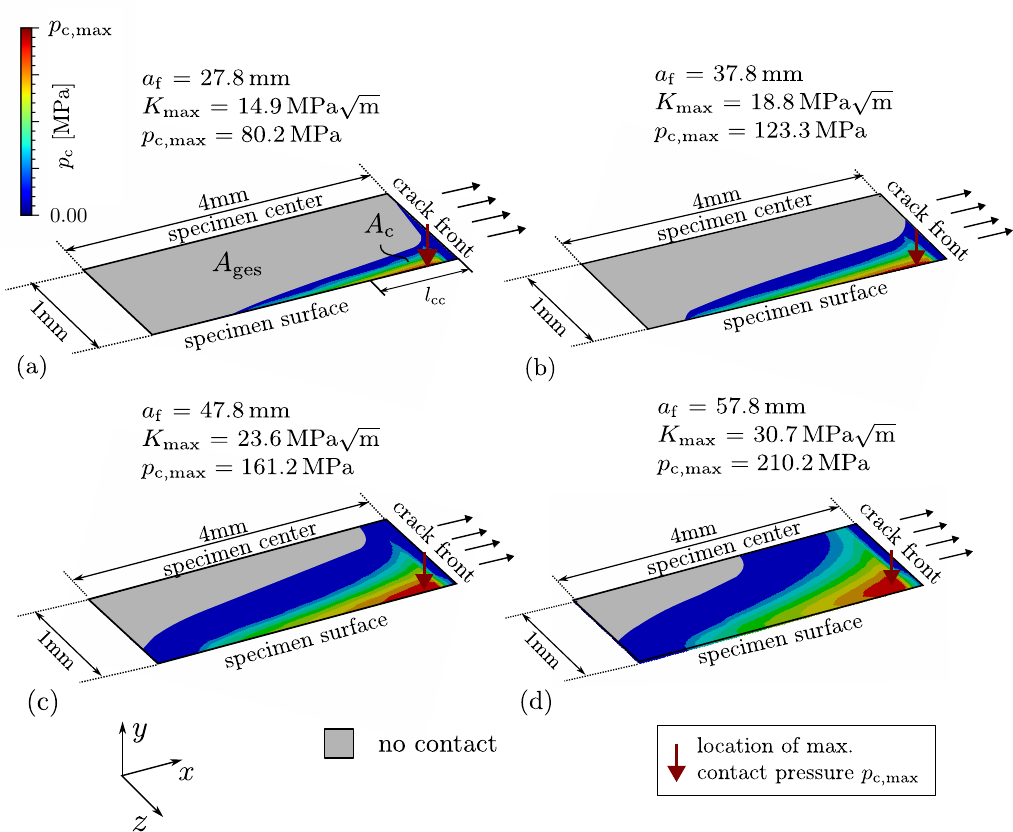}
	\caption{Contact pressure distributions at the crack surface obtained by the contact elements in the FE crack propagation model; The contact pressure distributions $p_\mathrm{c}$ are shown at four different crack lengths (a) $a_\mathrm{f}=\SI{27.8}{mm}$, (b) $a_\mathrm{f}=\SI{37.8}{mm}$, (c) $a_\mathrm{f}=\SI{47.8}{mm}$ and (d) $a_\mathrm{f}=\SI{57.8}{mm}$ at minimum load $F_\mathrm{min}=\SI{1.5}{kN}$}
	\label{fig:3d_picc}
\end{figure}

The FE and HR-DIC results agree on crack opening on the free surface, so we assume that PICC in the experiment behaves similarly to the FE simulation. Contact elements in the crack plane allow us examine the crack face contact through contact pressure distributions. Figure \ref{fig:3d_picc} shows the contact pressure  $p_{\mathrm{c}}$ on the crack surface as contour plots.

\begin{figure}[ht]
	\centering
	\includegraphics[width=0.95\textwidth]{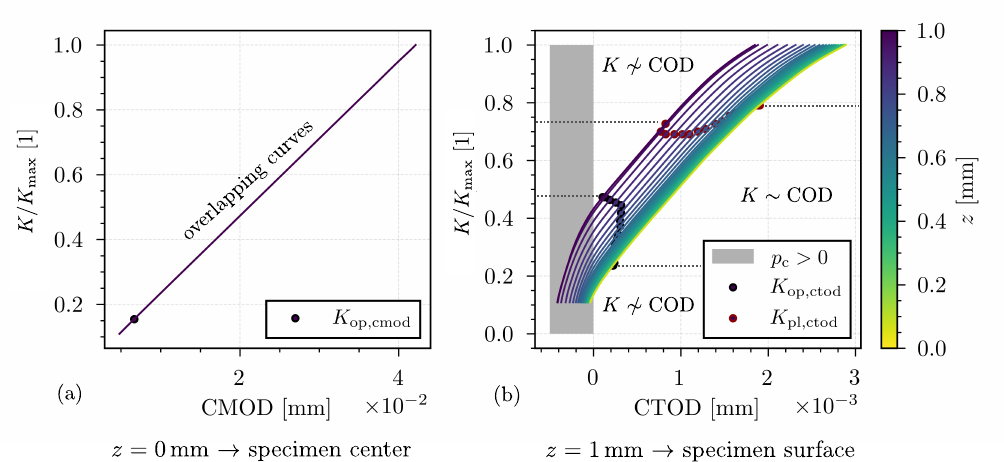}
	\caption{Comparison between the crack opening curves based on (a) CMOD and (b) CTOD measurements along the specimen thickness  $z$ ; values refer to FE crack propagation simulation data obtained at a crack length of $a_\mathrm{f}=\SI{27.8}{mm}$}
	\label{fig:cod_3d}
\end{figure}

All four images indicate that the contact pressure is directed towards the specimen's surface as a result of PICC, aligning with the assertions by \auth{Dugdale} \cite{Dugdale.1959} and \auth{Newmann} \cite{Newmann.1981}. They explain this to the significant influence of the plane stress state at the free surface of the specimen. Further FE-based studies have also identified that PICC is particularly prevalent under plane stress conditions \cite{McClung.1991, Calvin.2023, Calvin.2022, Kubicek.2022, Karkkainen.2023}. Moreover, Figure \ref{fig:3d_picc} shows that, due to the positive load ratio $R=0.1$, only a limited area directly behind the crack front stays in direct contact. Figure \ref{fig:3d_picc} clarifies that the contact pressure $p_{\mathrm{c}}$ increases with an increasing $K_{\mathrm{max}}$. The proportion of the contact area $A_{\mathrm{c}}$ to the total area $A_{\mathrm{ges}}$ also increases. At $a=\SI{57.8}{mm}$ the entire crack surface behind the crack front is in contact. Additionally, it is observed that the location of the maximum contact pressure $p_{\mathrm{c,max}}$ at $a_\mathrm{f}=\SI{47.8}{mm}$ shifts from the specimen's surface towards its center. This is particularly evident in Figure \ref{fig:3d_picc}d at a crack length of $a_\mathrm{f}=\SI{57.8}{mm}$. 
The variations in the contact pressure distributions can be attributed to the opposing effects of the plastic zone at the specimen's surface and the plane-strain in the specimen's center. As $K_{\mathrm{max}}$ increases, the plane stress state becomes more dominant, resulting in a higher proportion of the crack surface being in contact. Given that contact is primarily focused on the free surface, it is necessary to consider how the external contact of the crack faces influences the crack opening behaviour across the entire specimen thickness. This question is partially answered by Figure \ref{fig:cod_3d}. 
In order to characterise crack opening, the crack opening curves of both CMOD and CTOD nodes (the first row of nodes behind the crack front) along the specimen thickness are examined in Figure \ref{fig:cod_3d}. The basis for this investigation is the FE crack propagation simulation at $a=\SI{27.8}{mm}$ from Figure \ref{fig:3d_picc}.
Based on Figure \ref{fig:cod_3d}a, it is apparent that CMOD measurements, regardless of the measurement position in the thickness direction, lead to the same $K_{\mathrm{op,cmod}}$. Nevertheless, the algorithm-based evaluation (see Section \ref{sec:cod_method}) detects a $K_{\mathrm{op,cmod}} = \SI{2.53}{MPa\sqrt{m}}$, corresponding to $17\%$ of $K_{\mathrm{max}}$. The surface-near crack closure contact ($z=\SI{1}{mm}$) just behind the crack tip influences the crack opening behavior along the entire specimen thickness. The $K_{\mathrm{op,ctod}}$ values close to the specimen surface feature the largest crack opening values at $47\%$ of $K_{\mathrm{max}}$. Figure \ref{fig:cod_3d}b, besides illustrating the non-linearity due to crack closure, also shows another non-linear section starting from $70\%-80\%$ of the maximum load due to the large plastic strain accumulation in the plastic zone. 

This region is separated by $K_{\mathrm{pl}}$ as introduced in Section \ref{sec:cod_method} and is determined comparably to $K_{\mathrm{op}}$. \auths{Vasco-Olmo} \cite{VascoOlmoJ.M.2019} identify a similar non-linear behavior in their crack opening curves. They correlate it to the plastic strain accumulation in the plastic zone during crack opening. We support their hypothesis and extend it to three dimensions. The intermediate region between $K_{\mathrm{op,ctod}}$ and $K_{\mathrm{pl}}$ is characterized by linear-elastic crack opening without crack face contact.

\subsection{Effect of crack closure on plastic zone}
\begin{figure}[ht!]
	\centering
	\includegraphics[width=0.93\textwidth]{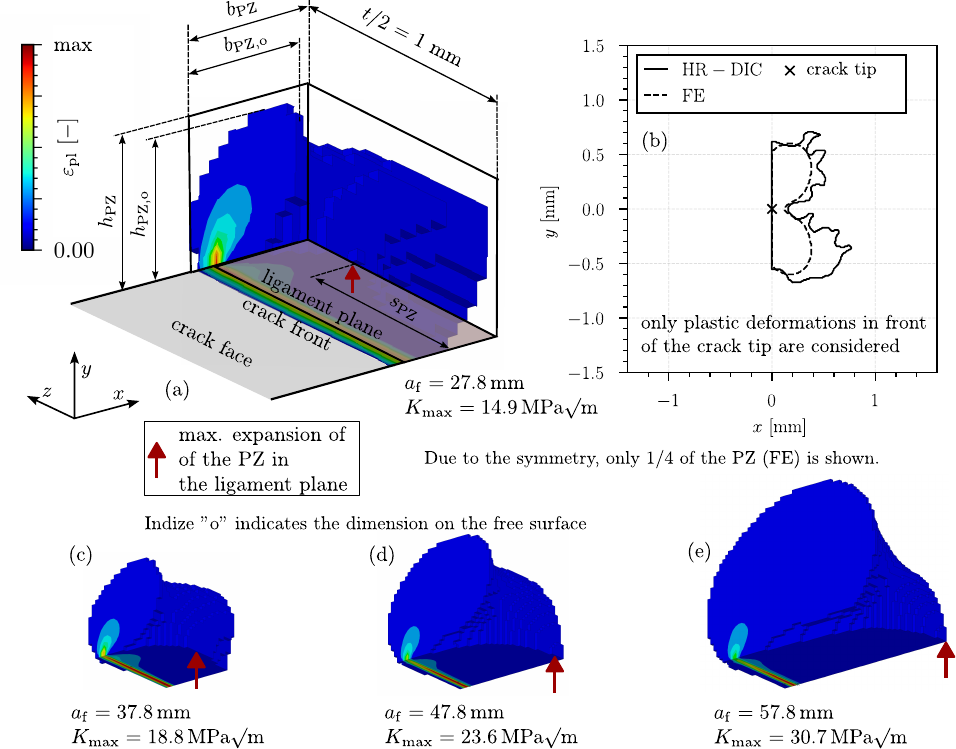}
	\caption{Shape analysis of the plastic zone: (a) 3D plastic zone at $a=\SI{27.8}{mm}$ including all investigated shape parameters, (b) 2D comparison of the plastic zone on the specimen surface based on FE and HR-DIC data and (c-e) 3D plastic zones at three further crack lengths}
	\label{fig:plastic_zone}
\end{figure}

In the following, we investigate the effect of the crack closure contact on the shape and damage evolution within the plastic zone. We consider as plastic zone (PZ) allplastic  deformations in front of the crack tip. Therefore, we select all elements with plastic deformation after the last opening load step in our FE simulations. The results are illustrated in Figure \ref{fig:plastic_zone}. Because of the symmetric FE model, only 1/4 of the entire PZ is displayed. Our results indicate that the PZ exhibits a shape that differs from the dog-bone model \cite{Dugdale.1959}. Similarly, the numerical shape investigations made by \auths{Camas} \cite{Camas.2012} and \auths{Besel} \cite{Besel.2016} are also in agreement with ours, although the effect of crack closure was not considered. Due to the extensive crack advancement, $\Delta a=\SI{0.08}{mm}$, employed in the FE simulation, it is only possible to reveal the primary PZ. Consequently, the cyclic PZ cannot be evaluated.
In order to analyse the impact of crack surface contact, we conducted FE crack propagation simulations with and without contact definitions. Subsequently, we compared the shapes of the resulting primary PZ. The use of elements with a cuboid shape facilitates quantitative assessments concerning the shape characteristics of the primary PZ. Figure \ref{fig:plastic_zone} shows the PZ shapes with active contact definition at four different crack lengths. Table \ref{tab:pz_shape} lists the results of selected descriptors for the shape of PZ. A summary about the parameter is illustrated in Figure \ref{fig:plastic_zone}a.
\begin{table}[h]
	\centering
	\begin{tabular}{|c|c|c|c|c|c|c|c|c|c|}
		\hline
		&& \multicolumn{4}{c|}{\textbf{without contact}} & \multicolumn{4}{c|}{\textbf{with contact}} \\
		
		 $a_\mathrm{f}$ & $K_\mathrm{max}$ & $h_\mathrm{PZ}$ & $b_\mathrm{PZ}$ & $s_\mathrm{PZ}$ & $V_\mathrm{PZ}$ & $h_\mathrm{PZ}$ & $b_\mathrm{PZ}$ & $s_\mathrm{PZ}$ & $V_\mathrm{PZ}$ \\
		 
		 [mm] & $\mathrm{[MPa\sqrt{m}]}$& $[\mathrm{mm}]$ & $[\mathrm{mm}]$ & $[\mathrm{mm}]$ & $[\mathrm{mm^3}]$ & $[\mathrm{mm}]$ & $[\mathrm{mm}]$ &$[\mathrm{mm}]$ & $[\mathrm{mm^3}]$ \\
		\hline
		27.8 & 14.9 & 0.38 & 0.34 & 1.68 & 0.24 & 0.38 & 0.34 & 1.68 & 0.22 \\
		37.8 & 18.8 & 0.76 & 0.80 & 1.24 & 0.46 & 0.76 & 0.80 & 1.24 & 0.45 \\
		47.8 & 23.6 & 1.42 & 1.79 & 0.24 & 0.92 & 1.42 & 1.79 & 0.24 & 0.90 \\
		57.8 & 30.7 & 1.58 & 2.49 & 0.00 & 1.32 & 1.54 & 2.45 & 0.00 & 1.30 \\
		\hline
	\end{tabular}
	\caption{Shape characteristics of plastic zone based on FE crack propagation simulation with and without contact definition.}
	\label{tab:pz_shape}
\end{table}
\begin{figure}[ht!]
	\centering
	\includegraphics[width=0.93\textwidth]{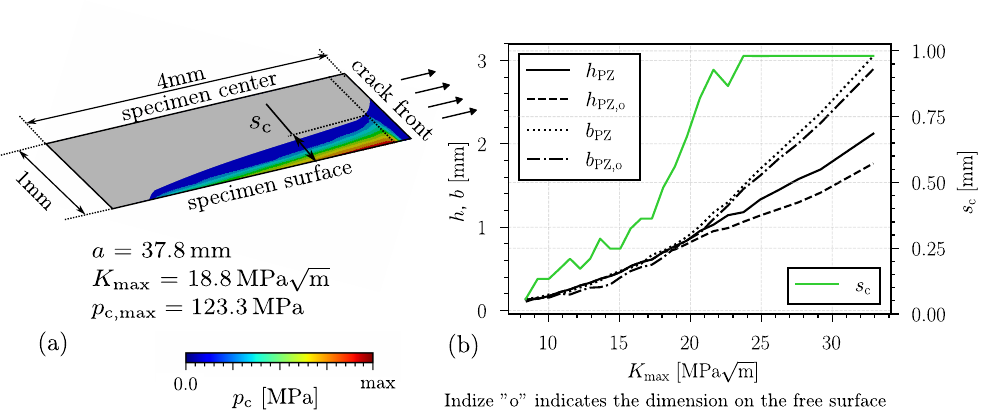}
	\caption{Correlation between PZ shape and crack surface contact situation: (a) exemplary illustration of a contact pressure distribution at $a_\mathrm{f}=\SI{37.8}{MPa\sqrt{m}}$ including the definition of $s_\mathrm{c}$ and (b) correlation between PZ height and width and the contact situation.}	\label{fig:pz_kontakt}
\end{figure}
Comparing the descriptors listed in the Table \ref{tab:pz_shape}, the deviations are all less than the size of one element, i.e. less than $2\,\%$ in all analysed parameters. Therefore, we conclude that the influence of the crack closure contact on the shape of the primary PZ is negligible. Whether the contact is confined to the crack surface near the free surface or encompasses the entire crack surface appears to be unimportant. Furthermore, the contact pressure distributions correlate with the shape of the primary plastic zone, as shown in Figure \ref{fig:pz_kontakt}. Figure \ref{fig:pz_kontakt}b plots the fraction of crack surface being in contact $s_\mathrm{c}$ to the entire specimen thickness (see Figure \ref{fig:pz_kontakt}a). The evaluation is based on the second row of elements behind the crack front to avoid numerical singularity effects that can result from the crack propagation algorithm. As illustrated in Figure \ref{fig:pz_kontakt}b, full-crack surface contact is present when the plastic zone is greater in width than in height. Since the shape of the plastic zone depends on both $K_\mathrm{max}$ and the specimen thickness $t$ \cite{Besel.2016}, this statement is so far only valid for the specimen under investigation ($t=\SI{2}{mm}$, AA2024-T3). Nevertheless, the impact of crack closure contact on the shape of the cyclic plastic zone remains to be investigated.

\begin{figure}[ht]
	\centering
	\includegraphics[width=0.9\textwidth]{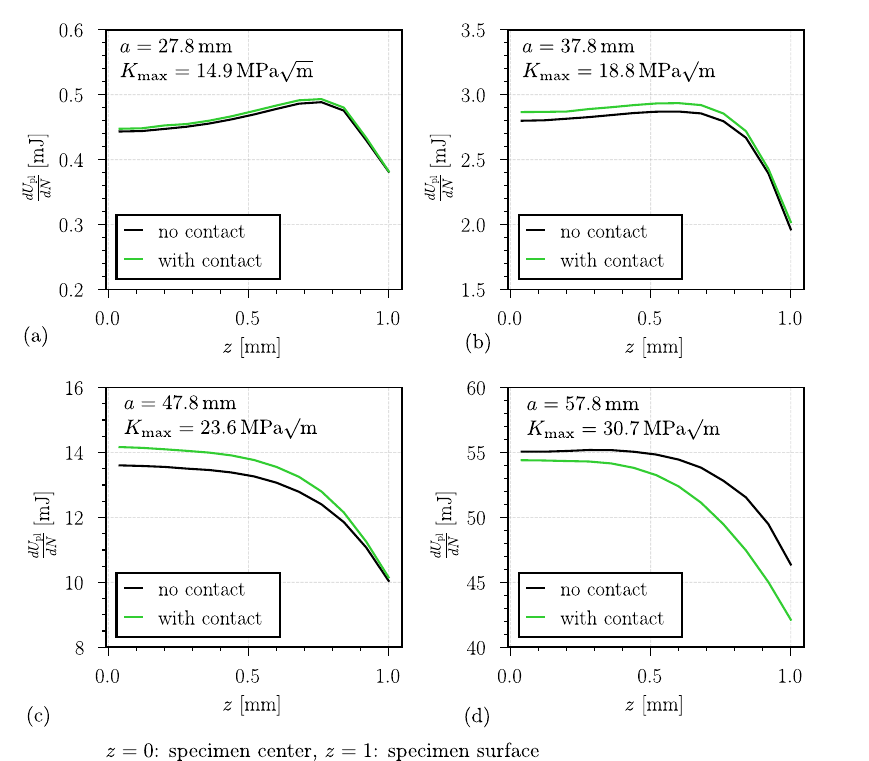}
	\caption{Effect of crack surface contact resulting from plasticity induced crack closure on the evolution of plastic energy within a single load cycle, evaluated at four different crack lengths: (a) $a_\mathrm{f}=\SI{27.8}{mm}$, (b) $a_\mathrm{f}=\SI{37.8}{mm}$, (c) $a_\mathrm{f}=\SI{47.8}{mm}$ and (d) $a_\mathrm{f}=\SI{57.8}{mm}$.}
	\label{fig:contact_damage}
\end{figure}

Figure \ref{fig:plastic_zone}b compares the shape of the plastic zone at the surface based on FE results with those based on the HR-DIC analysis at a crack length of $a=\SI{27.8}{mm}$. Both plastic zones are delineated from the linear-elastically deformed material by the yield strength $R_{\mathrm{p0.2}}=\SI{350}{MPa}$. It is observed that both shapes almost coincide indicating that the experimental results agree with the numerical findings. Furthermore, since the upper wing of the plastic zone is greater in height than in width, it indicates that the contact is only present near the free surface of the specimen. We neglect the lower wing because it has been found that its shape is strongly influenced by a HR-DIC measurement artifact resulting from low quality speckle pattern. Nevertheless, the plastic zone based on HR-DIC data is estimated to be slightly larger, attributed to the inherent measurement noise in HR-DIC data.
Because of the good agreement between the numerical and experimental results, we can analyse damage in the plastic zone due to plasticity induced crack closure. Therefore, we use the accumulation of plastic energy over a load cycle as damage describing parameter, following the suggestion by \auths{Vormwald} \cite{Vormwald.2016}. In our analysis, we compare the results of an FE simulation with contact definition to one without contact definition. This comparison helps to identify the influence of the contact on the energy development within the plastic zone. Figure \ref{fig:contact_damage} illustrates the distribution of plastic energy $\frac{d U_{\mathrm{pl}}}{d N}$ along the crack front $z$.

The two curves shown in  Figure \ref{fig:contact_damage}a deviate less than $1\%$, but $\frac{d U_{\mathrm{pl}}}{d N}$ with contact definition is slightly larger. This effect results from the numerical singularity that arises from the first FE element behind the crack tip and whose influence is slightly increased by the contact pressure. We observe the same trend for the crack lengths shown in Figures \ref{fig:contact_damage} b, c and d. The increased portion of the crack face being in contact combined with larger contact pressures increases the numerical singularity stress within the crack front element. This causes that the distance between the with contact and no contact curves increases. Only for larger crack lengths $a_\mathrm{f}=\SI{57.8}{mm}$ the contact pressure appears to reduce the accumulation of plastic energy per load cycle. The contact pressure is no longer focused towards the free specimen surface.  At $a=\SI{57.8}{mm}$ a plane stress state is pre-dominant for the entire plastic zone and the crack surface behind the crack front is in full contact. The findings indicate that when the contact pressure is concentrated on the free specimen surface, there is no notable impact on the plastic energy accumulation from PICC. However, when the maximum contact pressure point is shifted towards the specimen center and the entire crack face is in contact, PICC appears to exert a reducing effect on the plastic energy accumulation within the PZ.

\subsection{Fracture surface analysis}
\begin{figure}[ht]
	\centering
	\includegraphics[width=0.8\textwidth]{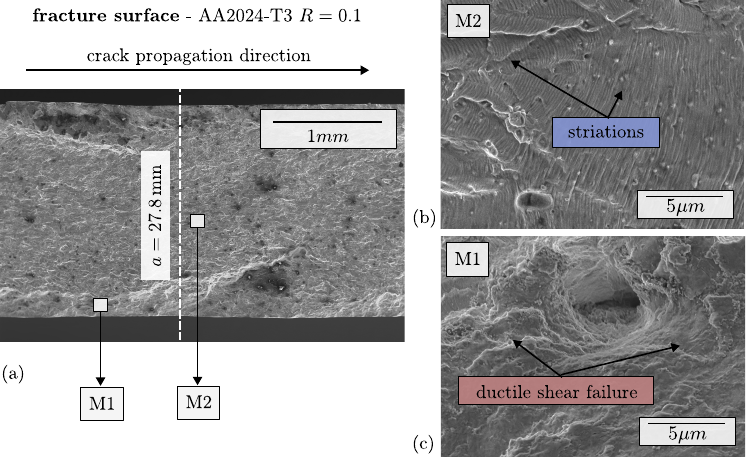}
	\caption{Fractographic analysis of the fracture surface near $a=\SI{27.8}{mm}$: (a) fracture surface on macroscopic scale, (b) striation mechanism and (c) ductile shear failure mechanisms near the free surface of the specimen}
	\label{fig:fracture_surface}
\end{figure}

Finally, we examined the fracture surface by scanning electron microscopy at different scales to find indications of crack surface contact near the surface. Figure \ref{fig:fracture_surface}a shows the fracture surface section whose displacement field is shown in Figure \ref{fig:teststand}c and to which all COD analyses in Figure \ref{fig:cod_paramstudie_exp} refer. The lower edge of the fracture represents the free specimen surface that has been investigated by the robot-supported HR-DIC measurement system. Before $a=\SI{27.8}{mm}$, two different damaging mechanisms can be identified within the fracture surface. In the center, as shown in Figure \ref{fig:fracture_surface}b, striations are observed. This aligns with the literature towards intrinsic fatigue crack growth mechanisms in AA2024 \cite{James.2003, Sakhalkar.2013}.

However, since the previous analysis has shown that the crack closure contact due to PICC is focused towards the crack surface edge, we study the intrinsic mechanisms in that particular region. Figure \ref{fig:fracture_surface}c shows that the fatigue crack propagates in that region by ductile shear failure mechanisms leading to a rough fracture surface close to \EEE the specimen surface. At $\Delta K=\SI{13.4}{MPa\sqrt{m}}$, fatigue crack growth in $t=\SI{2}{mm}$ thin MT(160) specimens made of AA2024-T3 tends to develop shear lips. It follows that the surface roughness in that region is increased. However, since the crack opening kinematics are still identical to  those observed in the FE crack propagation simulation, this indicates that the increased surface roughness due to the shear lips has a minor influence on the pre-dominant PICC mechanism in this section of AA2024-T3 fatigue cracks. That finding supports research performed by \auths{Materna} \cite{Materna.2019}. They concluded, based on numerical studies, that fracture surface roughness does not affect the crack closure behavior if PICC is the pre-dominant mechanism.

\section{Conclusions}
In this study, we investigate the crack closure behaviour of AA2024-T3 based on numerical simulations and experimental data. We find an excellent agreement between the crack opening curves based on FE and HR-DIC data. Given that the FE crack propagation model only considers the effect of PICC, we conclude that PICC is the dominant crack closure mechanism in AA2024-T3 fatigue cracks in our studies for an MT(160) specimen with $t=\SI{2}{mm}$ and a $\Delta K=\SI{13.4}{MPa\sqrt{m}}$. In addition, the agreements between simulation and experiment leads us to the following conclusions:

\begin{itemize}
	\item  PICC can be identified by using the $K_\mathrm{op,cod}$ dependence on the COD measurement location. Here, $K_\mathrm{op,cod}$ only depends on the horizontal distance to the crack tip position ($x_\mathrm{cod}$) and is independent of the vertical distance. However, we recommend to position the measurement points as close as possible to the crack path enabling a more stable identification of $K_\mathrm{op,cod}$.
	
	\item  Using the variation of $K_\mathrm{op,cod}$ values along the crack path, we introduce a new method for determining the opening value directly behind the crack tip $K_\mathrm{op,ctod}$. If PICC is present and the crack opening curves are shifted to a common minimum, the $K_\mathrm{op,cod}$ values form a linear relationship, that can be approximated by linear regression.
	
	\item The study supports the general assumption that PICC induces crack surface contact focused towards the free specimen surface. However, we show that even if there is only crack surface contact near the free surface of the specimen, it influences the crack opening behaviour throughout the specimen thickness.
	
	\item Considering a fatigue crack in AA2024 and a $t=\SI{2}{mm}$ thick MT(160) specimen, we found that the shape of plastic zone  is correlated to the crack surface contact situation. If the width of the plastic zone is larger than its height, this indicates that the entire crack surface close to the crack front is in direct contact. 
	
	\item Regarding the retardation effect of PICC on the damage accumulation within the plastic zone, we find  that the crack closure contact does not affect the plastic energy accumulation if the crack closure contact is directed towards the free surface of the specimen. If the entire crack surface is in contact, this seems to reduce the energy accumulation. 
	
	\item Based on fractographic investigations, we observe that fatigue cracks in AA2024-T3 at $\Delta K=\SI{13.4}{MPa\sqrt{m}}$ tend to develop shear lips. However, the increased surface roughness does not appear to affect the PICC mechanism. 
\end{itemize}

\section{Acknowledgements}

We acknowledge the financial support of the DLR-Directorate Aeronautics. This work was supported by the Deutsche Forschungsgemeinschaft, Germany (DFG) via the project Experimental analysis and phase-field modeling of the interaction between plastic zone and fatigue crack growth in ductile materials under complex loading (grant number BR 6259/2-1). Furthermore, funding came from the Federal Ministry for Economic Affairs and Climate Action, Germany on the basis of a decision by the German Bundestag, within the framework of the aerospace program LuFo-VI of the project "Intelligent FSW Process Monitoring" (Funding ID 20W2201E).

\section{Data availability}
The apdl finite element simulation code and data as well as the experimental displacement fields will be publicly available on Github (\url{https://github.com/dlr-wf}) and Zenodo (\url{10.5281/zenodo.13643861}).

\section{Competing interests}
The Authors declare no Competing Financial or Non-Financial Interests.

\section{Author contributions}
F.P. conceived the idea, conducted the simulations and experiments, evaluated the results, joined the discussions and wrote the manuscript. D.M. joined the discussions and wrote the manuscript. E.B. joined the discussions and wrote the manuscript.

\printbibliography

\end{document}